\documentclass[twoside,onecolumn,11pt,a4paper]{article}
\usepackage{amssymb}
\usepackage[latin1]{inputenc}
\usepackage{amsmath}
\usepackage{amsfonts}
\usepackage[margin=1.0in]{geometry}
\usepackage{graphicx}
\usepackage{caption}
\usepackage{subcaption}
\usepackage{sidecap}
\usepackage{cite}
\usepackage{authblk}
\begin{document}

\title{Trapped ion scaling with pulsed fast gates}
\date{}

\author[1]{C D B Bentley \thanks{christopher.bentley@anu.edu.au}}
\author[1,2]{A R R Carvalho}
\author[1]{J J Hope}

\affil[1]{Department of Quantum Science, Research School of Physics and Engineering, The Australian National University, Canberra, Australia}
\affil[2]{ARC Centre for Quantum Computation and Communication Technology, The Australian National University, Canberra, Australia}

\maketitle

\tableofcontents

\section{Abstract}
Fast entangling gates for trapped ion pairs offer vastly improved gate operation times relative to implemented gates, as well as approaches to trap scaling.
Gates on a neighbouring ion pair only involve local ions when performed sufficiently fast, and we find that even a fast gate between a pair of distant ions with few degrees of freedom restores all the motional modes given more stringent gate speed conditions. 
We compare pulsed fast gate schemes, defined by a timescale faster than the trap period, and find that our proposed scheme has less stringent requirements on laser repetition rate for achieving arbitrary gate time targets and infidelities well below $10^{-4}$.
By extending gate schemes to ion crystals, we explore the effect of ion number on gate fidelity for coupling two neighbouring ions in large crystals.
Inter-ion distance determines the gate time, and a factor of five increase in repetition rate, or correspondingly the laser power, reduces the infidelity by almost two orders of magnitude.
We also apply our fast gate scheme to entangle the first and last ions in a crystal.
As the number of ions in the crystal increases, significant increases in the laser power are required to provide the short gate times corresponding to fidelity above 0.99.

\section{Introduction} 

Trapped ion systems have performed quantum algorithms on small scales \cite{Guld03N,Chia04N,Chia05S,Bric05PRA,BlWi08N}, however achieving results inaccessible to the classical regime remains a significant challenge.
There are various entanglement protocols \cite{CZ95PRL,SoMo99PRL,SoMo00PRA,Milb00FP,CZ00N,Jona00PRA,Staa04PRA,Ospe08PRL,Roos08NJP}, many of which have been implemented \cite{Monr95PRL,Turc98PRL,Sack00N,DeMa02PRL,Leib03N,Schm03N,Haff05N,Leib05N,Halj05PRL,Halj05PRA,Home06NJP,Benh08NP,Kirch09NJP,Haye10PRL,Ospe11N,Monz11PRL,Berm12PRA}, where the required interaction between ions is mediated through the motion of the ions.
The quantised motional modes can be used as an information bus, where spin-state information can be mapped onto the motional state, and subsequently onto another ion's spin-state.
This idea was outlined in the original proposal for quantum information processing (QIP) using trapped ions \cite{CZ95PRL}.
Gates mapping information between spin and motional states must be performed in the regime where spectral resolution of sidebands can be performed.
This restricts the coupling strength to much less than the trap oscillation frequency, $\Omega \ll \nu$ \cite{Poya96PRA}.

QIP schemes that offer advantages over their classical counterparts require large ion and operation numbers: at least tens of qubits and thousands of operations for quantum simulations \cite{Lama14EPJ}.
The growing complexity of the motional spectrum with the number of trapped ions is a scaling problem for resolved-sideband gates.
The timescale of entangling gates increases relative to the trap frequency and decoherence timescales, which are on the order of seconds or minutes depending on the internal states \cite{Boll91IEEETIM,Haff08PR}.
The weak-coupling regime, where gate times with just two trapped ions are limited to much longer than the trap period, on the order of 1$\mu$s, is thus prohibitive for large-scale QIP operations in a single trap.
Scaling mechanisms using ion shuttling between traps and 2D architectures have been proposed, and promising steps taken in their implementation \cite{CZ00N,Kiel02N,Hens06APL,Home09S,Bowl12PRL,Walt12PRL,Berm13PRL}.
We explore the fast gate contribution as a scalable entangling operation within a single trap, which can be usefully combined with other scaling methods.

Fast gates were proposed by Garc\'ia-Ripoll, Zoller and Cirac in 2003 \cite{GZC03PRL} as a mechanical rather than spectral method for entangling ions. 
Here a controlled-phase gate is performed on two ions using state-dependent forces.
Instead of resolving particular motional sidebands, fast gates excite multiple motional sidebands in the strong coupling regime where $\Omega \gg \nu$, offering gate times no longer limited by the trap frequency.
The key element of an ideal fast gate is the motional control of the ions.
State-dependent forces drive the motion such that the spin-states acquire relative phase and are maximally entangled.

By moving into a new gate time regime, fast gates open new opportunities for scaling \cite{Duan04PRL, GZC05PRA, Zhu06EL,Zhu06PRL,Bent13NJP,Mizr13APB,Stea14NJP}.
The application of fast gates to two ions in a multiple-ion crystal was first outlined by Duan \cite{Duan04PRL}. 
Schemes faster than the local ion oscillation frequency are shown to entangle the ions, with arbitrary fidelity through multiple `cycles' of the fast gate scheme.
Instead of exciting collective modes of an ion crystal, which would leave electronic and motional states entangled at the end of a fast gate addressing just two qubits, \cite{Zhu06EL} shows that only local modes are involved in the gate.
More precisely, gates exceeding the local ion oscillation frequency only need consider the motion of neighbouring ions to those entangled by the gate.
This locality allows a small number of degrees of freedom to determine a high-fidelity two-qubit gate scheme addressing two ions in an arbitrary length ion crystal.
We explore this principle using pulsed fast gates, and extend fast gates to non-neighbouring ions.

In section~\ref{sec:formalism} we present the fast gate mechanism and conditions for the performance of a two-qubit gate, as well as the proposed pulsed gate schemes.
Recent progress has been made towards fast gates using high repetition-rate pulsed lasers \cite{Petr14OE, Mizr13PRL, Mizr13APB, Camp10PRL, Bent13NJP}, including implementation of the pulse pairs required for the gate \cite{Mizr13PRL, Camp10PRL} and an exploration of spin-motion entanglement in the strong coupling regime \cite{Mizr13APB}.
This highlights the need for the optimisation of implementable pulsed fast gate schemes, which we provide in section~\ref{sec:two}.
Here we compare the gate times and fidelities of the schemes given particular repetition rates, and provide the optimal regimes for each method.
Applying this knowledge of fast gates to ion crystals is the main thrust of this paper; in section~\ref{sec:scaling} we explore gate fidelities and laser requirements for entangling neighbouring ions in a long ion crystal.
We also present our results that a distant, non-neighbouring ion pair can be coupled in the fast regime using large momentum transfers, with motional restoration from simple symmetries in the pulse scheme.

\section{Fast gate formalism} \label{sec:formalism}

In this section we present a review of published fast gate schemes.

\subsection{Ideal gate and two-qubit conditions}
Fast gate schemes proposed with laser pulses use pairs of counter-propagating $\pi$-pulses to give state-dependent momentum kicks to the ions, exciting various motional modes.
The kicks are performed fast with respect to the ion trap dynamics.
While detuned pulses can be used to drive phase trajectories using Stark shifts as in \cite{Leib03N}, we focus on resonant transitions for simplicity.
These state-dependent trajectories correspond to state-dependent paths in phase space that provide a particular phase to the given state.
Area enclosed by a phase-space trajectory corresponds to the acquired phase \cite{Leib03N}.
We will consider particular phase-space trajectories, where we map the displacement and rotation of the centre of a coherent state. 
When the relative phase is $\pi/4$, we have performed the desired gate operation according to the ideal unitary:
\begin{align}
U_I = e^{i \frac{\pi}{4} \sigma_1^z \sigma_2^z} \Pi_{p=1}^L e^{-i \nu_p T_G a_p^\dagger a_p}. \label{eqn:idun}
\end{align}
Here $\sigma_k^z$ is the Pauli Z-operator on the $k$th ion, $\nu_p$ and $a_p$ are the frequency and annihilation operator for the $p$th motional mode, and $T_G$ is the total gate time.
Ions one and two are targeted by this gate operation, however in general the two target ions can be chosen as desired.

The product of rotation operators corresponds to the free evolution of the motional modes during the gate operation, over time $T_G$.  
There are $L$ modes corresponding to the number of ions in the trap.  
At the end of the ideal gate, the motional modes are restored, meaning that they have been rotated by the free evolution that would have occurred without a gate.
This restoration means that no heating is introduced by the operation.

As presented in~\cite{GZC03PRL,Bent13NJP}, the two-qubit ($L=2$) conditions for acquiring the desired phase and restoring the motional conditions are given by
\begin{align}
\Theta &= 4\eta^2 \sum^N_{m=2} \sum^{m-1}_{k=1} z_m z_k \left[ \sin(\nu \delta t_{mk}) - \frac{\sin(\sqrt{3}\nu\delta t_{mk})}{\sqrt{3}} \right] = \frac{\pi}{4} \label{Phase} \\
C_c &= \sum^N_{k=1} z_k e^{-i\nu t_k} = 0 \label{Cc}\\
C_r &= \sum^N_{k=1} z_k e^{-i \sqrt{3} \nu t_k} = 0, \label{Cr}
\end{align}
where $N$ groups of pulse pairs are applied with the $k^{\mathrm{th}}$ group of pairs containing $z_k$ pulses arriving at time $t_k$, and $\delta t_{mk} = t_m - t_k > 0$.
The sign of $z_k$ is the direction of the first pulse in each pair, and the trap centre of mass frequency and Lamb-Dicke parameter are $\nu$ and $\eta$ respectively. 
Setting $C_c$ and $C_r$ to 0 represents motional restoration and $\Theta = \pi/4$ represents the relative phase condition.
More general conditions for arbitrary numbers of ions in the crystal are derived in section~\ref{sec:genconds} in detail.

\subsection{Gate schemes}

Now that the conditions for an ideal fast gate operation have been established, we review proposed pulse schemes for applying the gate to a pair of ions.
A general optimization search for the optimal pulse scheme given laser parameters and complete pulse timing and direction freedoms is a prohibitively difficult problem.
Any restriction on the degrees of freedom for a more tractable search should ideally provide gate schemes with both an effective scaling of gate time with applied momentum, and a robust solution even for large numbers of ions.
The following schemes use the motional symmetries discussed in Appendix~1 for motional stability and a tractable search space.
We will show that each scheme has the optimal scaling of gate time with the momentum applied to the ions.

\subsubsection{GZC scheme} \label{sec:GZC}

The scheme proposed by Garc\'ia-Ripoll, Zoller and Cirac (GZC scheme)~\cite{GZC03PRL,GZC05PRA} is characterised by instantaneous groups of pulse pairs $\underbar{z}$ sent at times $\underbar{t}$, interspersed with free evolution:

\begin{center}
\begin{tabular}{ccccccc}
$\underbar{z}$ &= ($-2n$,	  &$3n$,  &$-2n$, &$2n$, &$-3n$, &$2n$) \\
$\underbar{t}$ &= ($-\tau_1$,  &$-\tau_2$,  &$-\tau_3$, &$\tau_3$, &$\tau_2$, &$\tau_1$).
\end{tabular} \\
\end{center} 

At time $-\tau_1$, $2n$ counter-propagating pulse pairs are applied along the trap axis (aligned with the $z$ axis) to provide a $4 n \hbar k$ momentum kick in the $-z$ direction.
The integer $n$ determines the gate time $T_G$, which scales with the total number of pulses in the scheme $N_p$ as $T_G \propto N_p^{-2/3}$.
For two ions this scheme exactly solves the condition equations~(\ref{Phase}),(\ref{Cc}) and (\ref{Cr}).

\begin{figure}[th]
     \centerline{\includegraphics[width=\columnwidth]{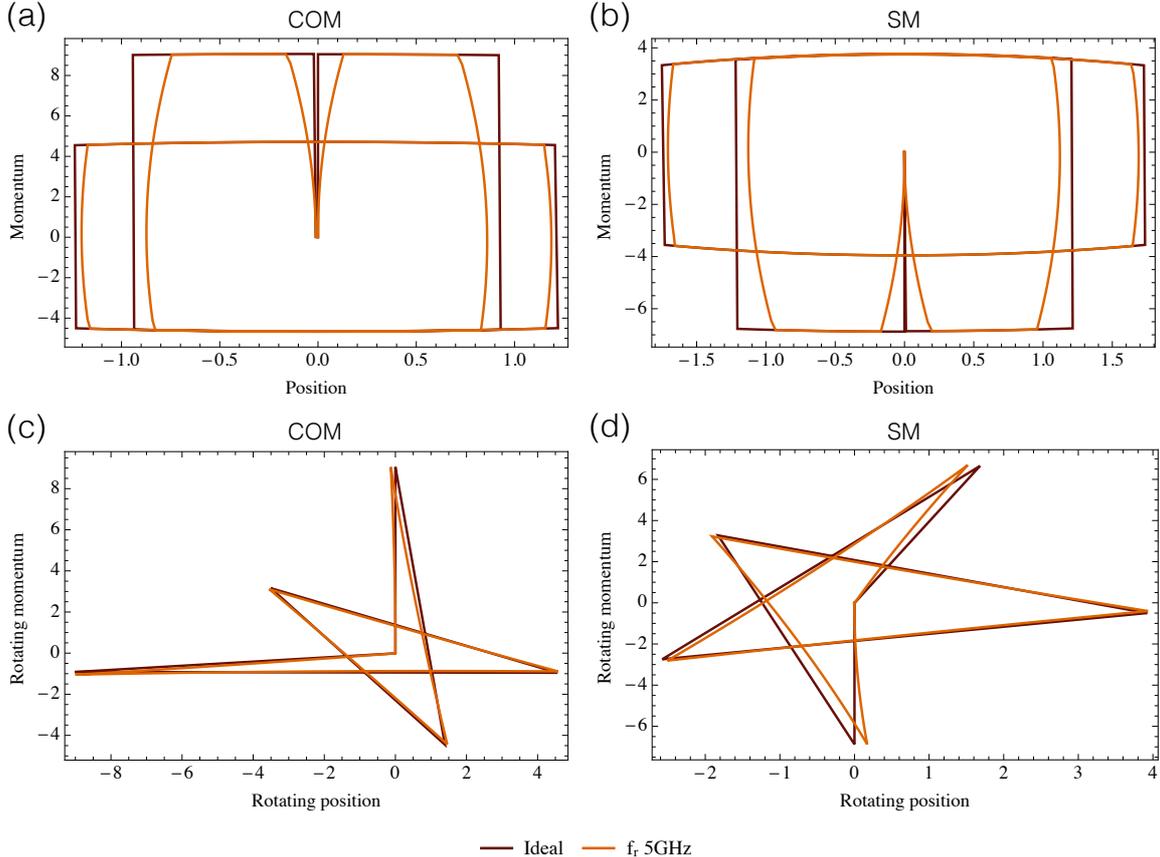}}
   \caption{GZC scheme in non-rotating (a,b) and rotating (c,d) phase space with $n=10$, for the centre of mass (a,c) and stretch (b,d) modes.
   Trajectories for the centre of a coherent state are plotted using dimensionless position and momentum.
   We show the performance of the ideal scheme, with an implied infinite repetition rate, and the performance of the scheme with a laser repetition rate $f_r = 5$GHz.
   For the centre of mass plots, both ions are in the excited internal state.  For the stretch mode plots, one ion is excited and the other is in the ground state.
   }
   \label{gzcphase}
\end{figure}

The unitary kick at a given time can be written as the product of unitary kicks on each of the modes.  
In Figure~\ref{gzcphase} we show the scheme applied to two ions with $n=10$ and a total gate time of $222$~ns, in rotating and non-rotating frames for both motional modes.
In rotating phase space, the initial motional state is stationary under no external evolution and the kick direction $i e^{-i 2 \pi \nu_p t_k}$ rotates depending on the time $t_k$ of the kick and the mode frequency.
Each mode-specific unitary has a kick strength that depends on the mode and the initial internal state of the ions.
The internal states determine the phase space trajectories: two ions sharing the same state are displaced in the centre of mass frame and invariant in the stretch mode, while the reverse is true for two ions with different internal states.  

Figure~\ref{gzcphase} also shows the effect of laser limitations, where instead of instantaneous pulses, each pulse pair is separated by the laser repetition period.
Instantaneous pulses at time $\tau_1$, for instance, become a group of $2n$ individual pulse pairs separated by the repetition period $\tau_r$ and centred in time at $\tau_1$.
For large numbers of pulses, or very short gate times, a limiting factor for the gate is avoiding temporal overlap for the different groups of pulse pairs separated by the finite repetition period.
As the laser repetition rate slows, the approximation to the instantaneous ideal pulses fails and the gate fidelity can drop significantly.
In this paper we focus on high repetition rates upwards of 300~MHz \cite{Petr14OE}, which we will show still provide remarkably high fidelities.

\subsubsection{Fast Robust Antisymmetric Gate scheme}

Using the antisymmetric form of the GZC gate scheme, a search for the optimal gate scheme for implementing fast gates by splitting pulses provided the following scheme \cite{Bent13NJP}, which we call the Fast Robust Antisymmetric Gate (FRAG):

\begin{center}
\begin{tabular}{ccccccc}
$\underbar{z}$ &= ($-n$, &$2n$, &$-2n$, &$2n$, &$-2n$, &$n$) \\
$\underbar{t}$ &= ($-\tau_1$, &$-\tau_2$, &$-\tau_3$, &$\tau_3$, &$\tau_2$, &$\tau_1$).
\end{tabular}
\end{center}

\begin{figure}[th]
     \centerline{\includegraphics[width=\columnwidth]{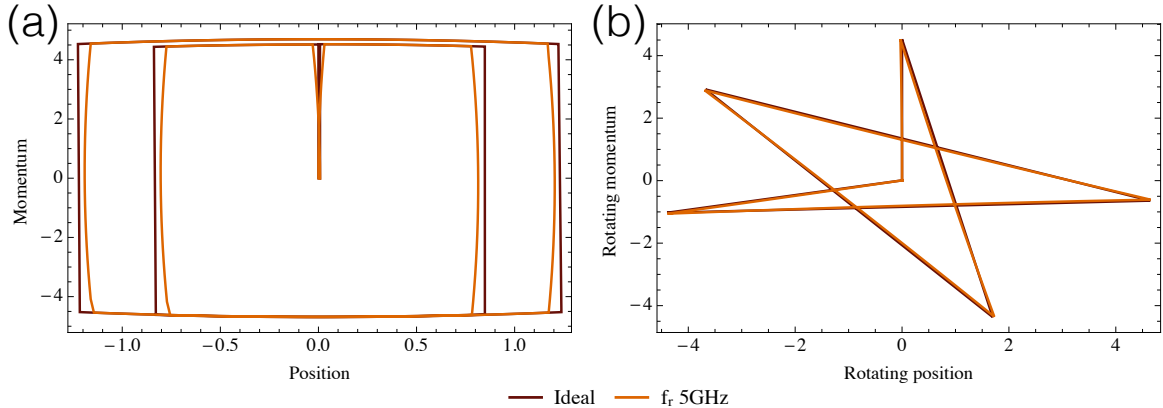}}
   \caption{FRAG scheme in non-rotating (a) and rotating (b) phase space with $n=10$, for the centre of mass mode with dimensionless position and momentum.
   The ideal scheme and one with a laser repetition rate of $f_r = 5$GHz are compared.
   Both ions are in the excited state.
   }
   \label{oursphase}
\end{figure}

Like the GZC scheme, it is an exact solution to the condition equations~(\ref{Phase}), (\ref{Cc}) and (\ref{Cr}) for two trapped ions, using the timing freedoms for motional stability.
Figure~\ref{oursphase} shows that the phase-space trajectories closely resemble those of the GZC scheme, and again the effect of a 5~GHz laser repetition rate is shown.
This scheme was designed for splitting large-area pulses, but we will explore other advantages of this scheme for $\pi$-pulse pairs directly applied to the ions.

\subsubsection{Duan scheme}

A simplified version of the scheme presented in \cite{Duan04PRL}, is described by

\begin{figure}[th]
     \centerline{\includegraphics[width=\columnwidth]{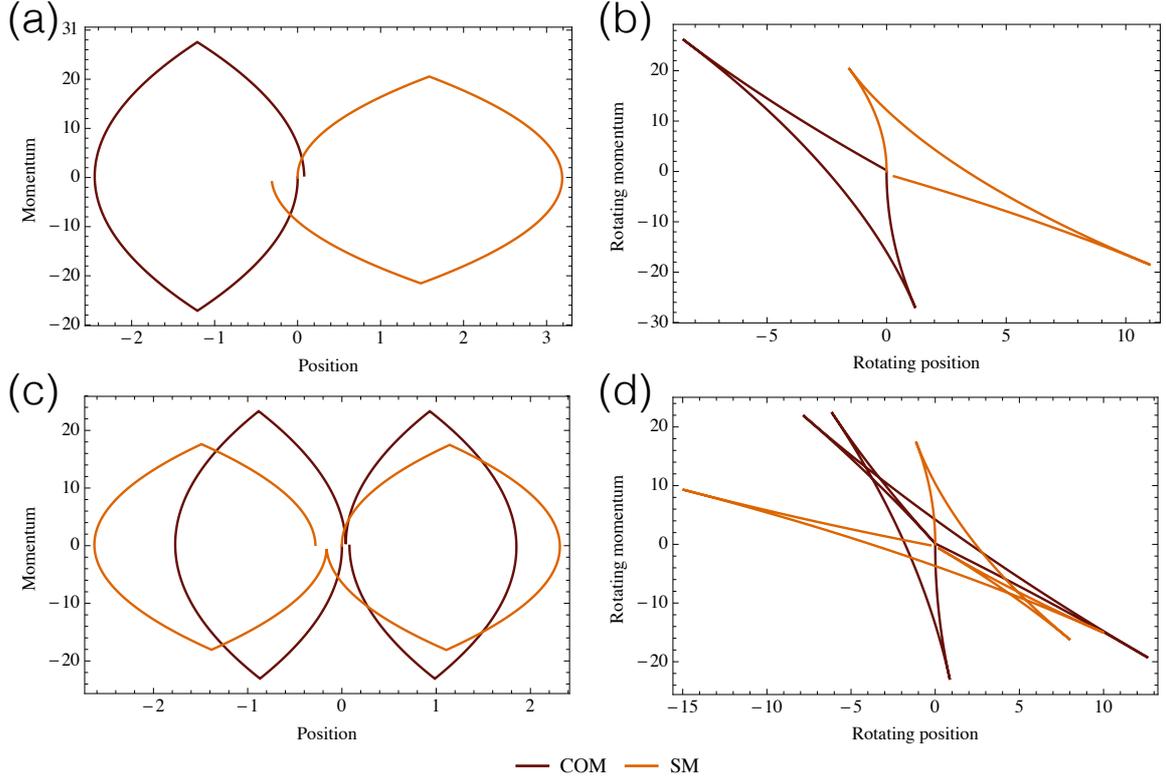}}
   \caption{Duan scheme in non-rotating (a,c) and rotating (b,d) phase space with $n=10$, for a single-cycle (a,b) and double-cycle (c,d) scheme.
   The centre of mass (COM) and stretch mode (SM) trajectories are shown.
   For the COM plots, the ions are in their excited states, while for the SM plots one ion is excited and one in the ground state.
   The laser repetition rate is $f_r = 5$GHz, which provides the phase space curvature of the scheme.
   For the single-cycle scheme, $\tau_1 = 60 \tau_r$, while for the double-cycle scheme $\tau_1 = 51 \tau_r$ such that the phase gate is performed.
   Both schemes are not quite restored to their initial motional state, most clearly seen in (b) and (d), however the double-cycle scheme in (d) is more robust.
   }
   \label{duanphase}
\end{figure}

\begin{center}
\begin{tabular}{cccc}
$\underbar{z}$ &= (n, &-2n, &n) \\
$\underbar{t}$ &= (0, &$\tau_1$, &$2\tau_1$),
\end{tabular}
\end{center}
where we have assimilated pulses with the same direction into instantaneous kicks to match the form of GZC and FRAG.
As presented in \cite{Duan04PRL}, the scheme was designed for pulsed lasers with finite repetition period $\tau_r$ such that each pulse pair is separated by this period:

\begin{center}
\begin{tabular}{cccccccccccc}
$\underbar{z}$ &= (1, &1, &1, &..., &1, &-1, &..., &-1, &1, &..., &1) \\
$\underbar{t}$ &= (0, &$\tau_r$, &$2\tau_r$, &..., &$\tau_1-\tau_r$, &$\tau_1$, &..., &$3\tau_1 - \tau_r$, &$3\tau_1$, &..., &$4\tau_1$).
\end{tabular}
\end{center}
The scheme was designed for minimal gate time, such that the repetition period between pulses is the only free evolution, and the scheme forms a triangle-like shape in rotating phase-space, as shown in Figure~\ref{duanphase}.

This scheme takes the simplest form of a symmetry discussed in Appendix~1 to complete the triangular loop, using a single phase-space loop for maximal area rather than multiple triangular shapes as in the star-shaped schemes we have seen.
Doubling the scheme, so that there are two phase-space cycles, cancels the first order of motional error.
Here the scheme is simply repeated such that every pulse pair gives the opposite direction momentum kick to those in the first cycle.
Figure~\ref{duanphase} shows the doubled scheme, which takes almost twice as long as the single cycle for the 5~GHz repetition rate.

Even with multiple cycles, the Duan scheme involves only one free variable in time, which is chosen such that the phase equation is satisfied.
The scheme is thus not an exact solution to the condition equations.
Accordingly, the symmetries employed provide gate speed requirements such that the error terms are suitably small for motional stability.
Multiple loops or cycles, appropriately designed, increase the motional stability as found in Appendix~1.
We will construct the scheme with different numbers of cycles to enclose the approximate desired phase, then optimise to explore the tradeoff between fast phase acquisition and fidelity.

\section{Two trapped ions: gate optimisation and application} \label{sec:two}

The three presented schemes provide solutions to the fast gate condition equations~(\ref{Phase}), (\ref{Cc}) and (\ref{Cr}) for two ions in a trap.
We use simulated annealing to optimise the pulse timings and the number of pulses $n$ to maximise the gate fidelity and minimise the gate duration for a given repetition rate.
We compare the resulting gate times and fidelities for each scheme.

\subsection{Fidelity measure}

Our performance measure is the state-averaged fidelity, derived in Appendix~2, which depends on the number of ions.
For two ions,
\begin{align}
F_{2\text{ave}} = \frac{1}{12} \left(6 + e^{-4 m |x|^2} + e^{-4 m |y|^2} + 4 e^{-m (|x|^2 + |y|^2)} \cos (\phi') \right), \label{eq:fidmeas}
\end{align}
where $x$ and $y$ are the displacements in phase space for each mode, $\phi'$ is the actual relative phase difference minus the desired $\pi/2$ for different initial states.
There is also dependence on the initial motional state, since
\begin{align}
m \equiv (\frac{1}{2}+\bar{n}),
\end{align}
for the mean mode occupation $\bar{n}$, which is set to be equal for each mode.
We will vary the initial mode occupation $\bar{n}$, and explore the effect of trap temperature on fidelity, in section~\ref{sec:2sch}. 

The fidelity measure takes values between 0.5 and 1 due to the state-averaging effect, for relative phase between 0 and $\pi$.
For different numbers of ions, the fidelity function becomes more complex.
However, it has the same exponential dependence on the initial motional state and the final phase space trajectory displacement from the ideal, as well as the sinusoidal phase dependence.

\subsection{Scheme performance for $^{40}$Ca$^+$} \label{sec:2sch}

We use the $^{40}$Ca$^+$ ion for our fast gate analysis as a typical ion candidate.
Our parameters are
\begin{align}
\nu &= 2 \pi \times 1.2 \text{ MHz}\\
\lambda &= 393 \text{ nm}\\
\eta &= 0.16 \\
\bar{n}_1 &= \bar{n}_2 = 0.1 \text{ phonons},
\end{align}
where we use the $S_{1/2}$ to $P_{3/2}$ transition to give the ions momentum kicks.
The computational excited state, the metastable $D_{5/2}$ state, is untouched by the resulting momentum kicks.
Since the forces are only applied to the computational ground $S_{1/2}$ state, this differs from the above derivation where equal and opposite kicks are assumed for the two states.
The adjustment results in an effective halving of the momentum kick size in the condition equations derived, as shown in Appendix~3.

For the given ion parameters, Figure~\ref{Timeforce} shows the gate time as a function of laser repetition rate, defined here as the rate at which a pair of counter-propagating $\pi$-pulses can be applied to the ions.
Higher repetition rates correspond to stronger forces displacing the ions from their equilibrium positions, and the relative phase is acquired more quickly.
Low pulse numbers in a scheme correspond to slower gates, such as the $n=5$ (70 pulse pair) GZC scheme and the $n=9$ (90 pulse pair) FRAG scheme for a 300~MHz repetition rate.
Each scheme in the Figure has the optimal scaling of the repetition rate $T_G \propto f_r^{-2/5}$ as the repetition rate increases.
This scaling is from the optimal gate time with total pulse number relationship \cite{GZC05PRA}
\begin{align}
T_G \propto N_p^{-2/3},
\end{align}
with the approximation that $N_p = T_G f_r$.

\begin{figure}[th]
     \centerline{\includegraphics[width=0.7\columnwidth]{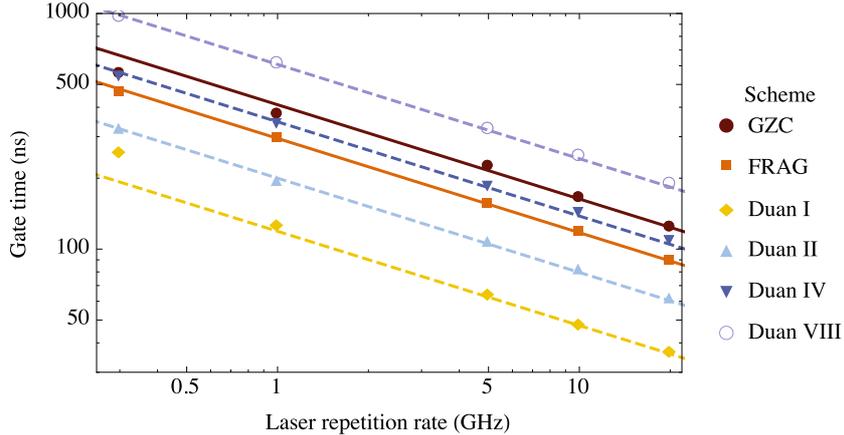}} 
   \caption{Gate time as a function of repetition rate for the presented schemes.  
   The fits show the optimal scaling of gate time with pulse number.
   Solid lines are fit to the high-fidelity FRAG and GZC gates with error ($1-$Fidelity) below $10^{-8}$.
   Dashed lines are fit to the Duan schemes, which have much lower fidelity.  
   The Roman numerals enumerate the number of cycles, or triangles, in the Duan scheme.  
   }
   \label{Timeforce}
\end{figure}

\begin{figure}[th]
     \centerline{\includegraphics[width=0.7\columnwidth]{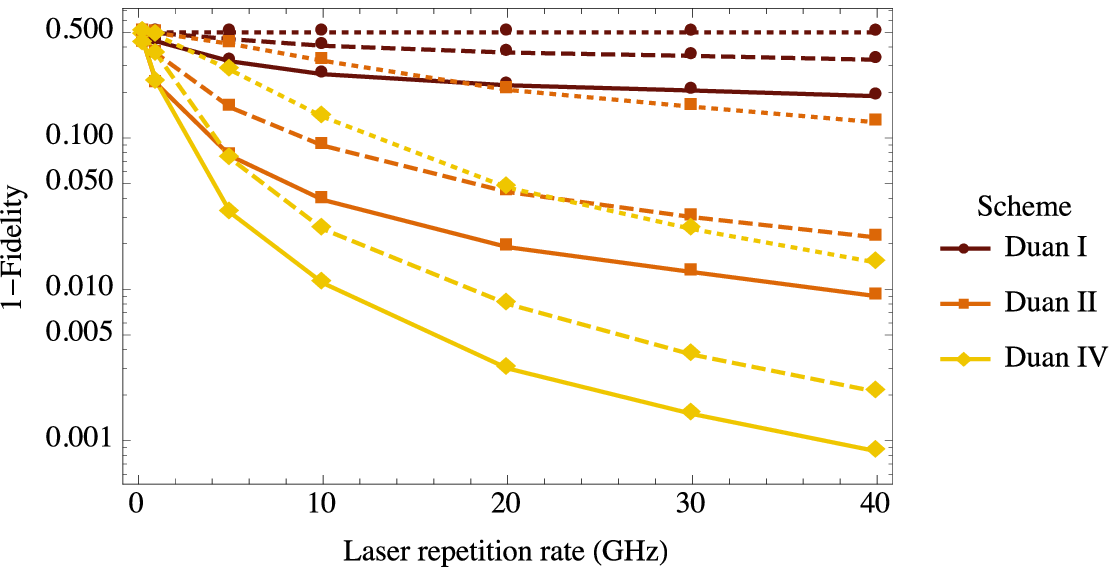}} 
   \caption{Gate error (1-Fidelity) as a function of repetition rate for the Duan cycle schemes.  
   The roman numerals mark the number of cycles (triangles) in the plotted scheme. 
   More cycles reduce the error for a given repetition rate, and increasing the repetition rate increases the motional robustness of the scheme.
   The solid line is for mean initial motional mode occupation $\bar{n}_1 = \bar{n}_2 = 0.1$, while the dashed and dotted lines represent mode occupations of 1 and 10 respectively.
   }
   \label{Fidforce}
\end{figure}

The high-fidelity schemes (FRAG and GZC) and the Duan schemes are marked by solid lines and dashed lines respectively in Figure~\ref{Timeforce}.
The high-fidelity schemes have error ($1-$Fidelity), or infidelity, below $10^{-8}$, while the Duan schemes have much lower fidelity as shown in Figure~\ref{Fidforce}.
The number of cycles of the Duan scheme is marked by Roman numerals in Figures~\ref{Timeforce} and \ref{Fidforce}.

Faster repetition rates lead to higher gate fidelities, as shown in Figure~\ref{Fidforce}.
The faster gate times simplify the motional conditions, as discussed in section~\ref{sec:gscal}.
More Duan cycles also increase the robustness of the scheme to motional error.
However, the gate time for four or more cycles of the Duan scheme was shown to be longer than the FRAG scheme.

High motional occupation enhances the error according to the exponential terms in the fidelity function, up to the maximal error asymptote at 0.5.  
For very high fidelity gates, such as the two-ion GZC scheme with finite repetition rates, the infidelities for mode occupations of 1 and 10 were found to remain on the order of $10^{-8}$.  

We will focus on the GZC and FRAG gates, with very high fidelities, as we extend the fast gate schemes to target two ions in ion crystals of varying length.
The Duan schemes with one or two cycles may be faster, but have insufficient fidelity for QIP protocols with foreseeable laser repetition rates.

\section{Scaling ion number: gate optimisation and application} \label{sec:scaling}

Here we consider longer ion crystals, $L>2$, and the efficacy of applying entangling gates to both a neighbouring and distant ion pair.
We thus explore the impact of fast gates on the scaling problem for trapped ions.

\subsection{General gate conditions} \label{sec:genconds}

We can extend the two-ion gate conditions to more general conditions for ion crystals with arbitrary length.
Fast gate evolution is composed of momentum kicks and the free evolution of the ion motion.
The momentum kicks are assumed to be fast relative to the ion motion, and evolve the state according to
\begin{align}
U_{kick} = e^{-2izk(x_1 \sigma_1^z + x_2 \sigma_2^z)} \label{eq:ukick}
\end{align}
for a pair of resonant $\pi$-pulses addressing ions 1 and 2, chosen arbitrarily.    
Here $z$ is the number of counter-propagating $\pi$-pulse pairs comprising the momentum kick, $k$ is the laser wavenumber, $x_1$ and $x_2$ are the ion positions.
We consider pulse lengths much longer than atomic transition periods $\sim 10^{-15}$~s to satisfy the rotating wave approximation.

The ion position can be written in terms of mode displacements and ion-mode coupling coefficients $b_i^{(p)}$ \cite{Jame98APB},
\begin{align}
x_i &= \sum_p^L b_i^{(p)} Q_p, 
\end{align}
where the dimensional mode position operator $Q_p$ is given by the mode annihilation and creation operators:
\begin{align}
Q_p  = \frac{\eta_p}{k} (a_p + a_p^\dagger).  
\end{align}
Using the Lamb-Dicke parameter,
\begin{align}
\eta_p = k \sqrt{\frac{\hbar}{2 M \nu_p}},
\end{align}
the momentum kick is composed of displacement operators for each motional mode $p$:
\begin{align}
U_{kick} &= \Pi_{p=1}^L  e^{-2iz (b_1^{(p)} \sigma_1^z + b_2^{(p)} \sigma_2^z) \eta_p (a_p + a_p^\dagger)} \\
              &= \Pi_{p=1}^L  \hat{D}_p (-2iz (b_1^{(p)} \sigma_1^z + b_2^{(p)} \sigma_2^z) \eta_p).
\end{align}
where there are $L$ modes for $L$ ions in the chain, and the displacement operator,
\begin{align}
\hat{D}_p(\alpha) = \exp [ \alpha a_p^\dagger - \alpha^* a_p ],
\end{align}
commutes for each mode $p$.

The free motional ion evolution corresponds to a phase space rotation for each mode:
\begin{align}
U_{p,mot} = e^{-i \nu_p \delta t_k a_p^\dagger a_p}
\end{align}
where $\delta t_k$ is the time between the $k$th and $(k+1)$th momentum kicks.

The total gate evolution is described by momentum kicks with direction $z_k$ interspersed with free evolution until the next momentum kick:
\begin{align}
U_{gate} &= \Pi_{k=1}^N \Pi_{p=1}^L \hat{D}_p (-i c_{pk}) e^{-i \nu_p \delta t_k a_p^\dagger a_p}, \\
c_{pk} &\equiv 2 z_k (b_1^{(p)} \sigma_1^z + b_2^{(p)} \sigma_2^z) \eta_p.
\end{align}
where $N$ is the total number of pulse pairs.

For a given mode $p$, the products of displacements and rotations give a total unitary 
\begin{align}
\mathcal{U}_p |\alpha \rangle = e^{i \xi_p} |\tilde{\alpha} \rangle,
\end{align}
acting on an initial coherent state $|\alpha \rangle$.  Here
\begin{align}
\tilde{\alpha} &= \alpha e^{-i \nu_p T_G} -i \sum^N_{k=1} c_{pk} e^{i \nu_p (t_k - T_G)}, \\  
\xi_p &= \sum^N_{m=2} \sum^{m-1}_{k=1} [c_{pm} c_{pk} \sin (\nu_p (t_m - t_k))] - \text{Re} \left[ \alpha \sum_{k=1}^N c_{pk} e^{-i \nu_p t_k}  \right]. \label{eq:fullcohphase}
\end{align}
The internal state is left invariant.

The state $|\alpha\rangle$ has been displaced in motional phase space, then rotated according to the ideal unitary in equation (\ref{eqn:idun}).
The phase acquired is caused by the ion excursion from equilibrium and subsequent different potential experienced, giving rise to phase evolution terms.

The total displacement is by an internal state-dependent amount  
\begin{align}
C_p = -i \sum^N_{k=1} c_{pk} e^{i \nu_p t_k}, \label{eq:pmotcondfull}
\end{align}
which we desire to be zero for the ideal gate process.  This displacement also provides the unwanted phase term,
\begin{align}
\xi_p' = e^{-i \text{Re} \left[ \alpha \sum_{k=1}^N c_{pk} e^{-i \nu_p t_k}  \right]},
\end{align}
which does not provide relative phase conditional on the states of two ions, as in (\ref{eqn:idun}).  
It is state dependent, and thus gives a relative phase term for each mode.
When the unwanted phase-space displacement is zero, this relative phase is also zero.
For nonzero displacements, the single-ion phase terms can be cancelled with single qubit rotations if significant.

The condition for ideal motional evolution for each mode is thus
\begin{align}
0 &=  -i \sum^{N}_{k=1} c_{pk} e^{i \nu_p t_k}  \\ 
 &= \sum^N_{k=1} z_k e^{i \nu_p t_k}. \label{eq:pmotcond}
\end{align}

We expand the remaining phase term for mode $p$, neglecting global phase terms without internal state dependence:
\begin{align}
\xi_p &= \sum^N_{m=2} \sum^{m-1}_{k=1} c_{pm} c_{pk} \sin (\nu_p (t_m - t_k)) \\
	 &= 8 \eta_p^2 \sigma_1^z \sigma_2^z b_1^{(p)} b_2^{(p)} \sum^N_{m=2} \sum^{m-1}_{k=1} z_m z_k  \sin (\nu_p (t_m - t_k)). \label{eq:redcohphase}
\end{align}

The phase for $L$ ions is given by the product of the phase from each mode,
\begin{align}
\Pi^L_{p=1} e^{i \xi_p},
\end{align}
thus to obtain the ideal unitary in equation (\ref{eqn:idun}), we need
\begin{align}
\frac{\pi}{4} = 8 \sum^L_{p=1} \eta_p^2 b_1^{(p)} b_2^{(p)} \sum^N_{m=2} \sum^{m-1}_{k=1} z_m z_k \sin(\nu_p(t_m-t_k)). \label{eq:pphasecond}
\end{align}

Equations~(\ref{eq:pmotcond}) and~(\ref{eq:pphasecond}) provide $L+1$ conditions to perform a fast gate with $L$ ions.
Note that for two ions, our derivation follows that of \cite{GZC03PRL}.

\subsection{Gate scaling} \label{sec:gscal}

Entangling gates that require resolution of particular motional sidebands become increasingly challenging as the mode density increases, which occurs as the number of ions in the trap goes up.
Since the gate time must be much longer than the frequency splitting \cite{Choi14PRL}, this causes these gates to slow dramatically.
In contrast, fast gates do not couple to particular sidebands, however the number of motional conditions for the gate are proportional to the number of modes, or ions, in the trap.
Solving these equations independently becomes prohibitively complicated for a practical gate, which would ideally be independent of the number of ions.

The fast gate timescale provides a solution to this problem.
A gate for two neighbouring ions that is much faster than the local oscillation frequency of other ions in the crystal only needs to satisfy the condition equations for local ions.
This idea is applied in \cite{Zhu06EL}, using amplitude-controlled segments from a continuous-wave laser to demonstrate the reduction in degrees of freedom required for sufficiently fast gates on neighbouring ions.
In fact, as we demonstrate in Appendix~1, only two equations are required for suitably stable and fast pulse schemes, and motional symmetries of different forms protect against different orders of the motional error.
A given gate scheme is motionally stable below a gate time that depends on the degrees of freedom and symmetries of the scheme.
This scheme-dependent stability is evident for two ions in section~\ref{sec:two}.
Faster gate times push each scheme towards motional stability.
Faster gates also accomodate more ions in the crystal, with higher local oscillation frequencies of the ions.

Performing gates on non-neighbouring ions, in contrast to the locality exploited above, will necessarily involve the motion of the intermediary ions.
However, the same reduced motional conditions of Appendix~1 restore the motion of each ion, thus there is a regime where the gate can be performed robustly.
In deriving the local oscillation frequency, each ion is assumed to be at equilibrium, and here we require momentum kicks sufficiently large to break this assumption and couple ions within a timescale shorter than the oscillation frequency.

Gates coupling non-neighbouring or distant ions in a chain via optimal control of laser parameters have been proposed, following the fast gate formalism.
In \cite{Zhu06PRL,Lin09EPL}, the ion coupling is mediated by transverse phonon modes using small numbers of continuous pulse segments, resulting in gates much longer than the trap period.
We will explore the pulsed laser requirements to perform gates between non-neighbouring ions with high fidelity and speed, as well as the impact on the intermediate ions.

Two ions can be addressed in long crystals using lasers that couple the ions through transverse modes, and it is straightforward to adapt our formalism to this case.  We consider instead the axial modes, assuming that ions not targeted by the gate can have the addressed internal states shelved in non-interacting states to allow the lasers to shine on or close to the crystal axis.   

\begin{figure}[t]
     \centerline{\includegraphics[width=0.6\columnwidth]{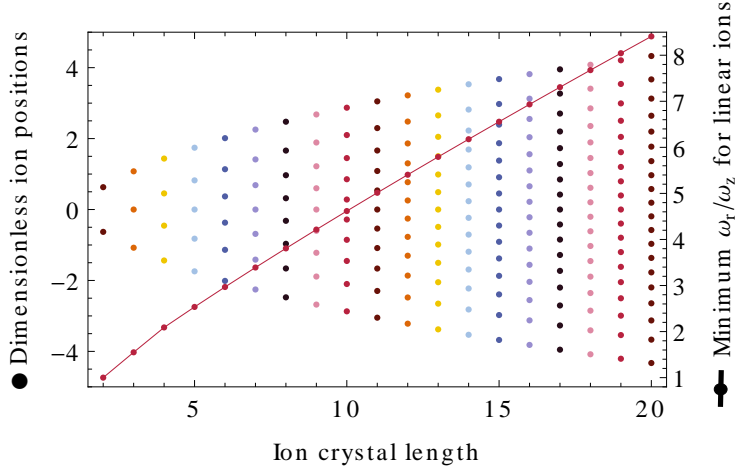}} 
   \caption{Equilibrium ion positions are shown in dimensionless units for different crystal lengths.
   For linear confinement of $L$ ions, the ratio between radial and axial COM frequencies $\omega_r/\omega_z > 0.63 L^{0.865}$ \cite{Wine98JRNIST,Schi93PRL}, and we plot the lower bound equality with the crystal length.
    }
    \label{fig:eqpos}
\end{figure}

Within a single trap, different gate schemes are required depending on the chosen ion pair.
Coulomb repulsion leads to tighter grouping of the ions in the centre of the trap, shown in Figure~\ref{fig:eqpos}, which leads to varying inter-ion coupling strengths between neighbouring ion pairs.
Alternatively, an anharmonic trap such as in \cite{Lin09EPL} has constant inter-ion distances, which would prove useful for coupling different ions with the same gate scheme instead of adapting for varying distance.

It is important to note that even with perfectly scalable entangling gates, a single trap cannot be scaled to include an arbitrary number of ions.
A strong radial frequency confinement is required relative to the axial frequency confinement so that the ions are prevented from buckling to a zig-zag formation, as shown in Figure~\ref{fig:eqpos}.
Lowering the axial frequency and addressing the radial modes has been used to manipulate large numbers of ions in a single trap \cite{Jurc14N}.
The anharmonic trap in \cite{Lin09EPL} provides stable confinement for many ions, and fast gates have also been proposed for a 2D architecture using controllable continuous pulse segments \cite{Zou10PLA}.
While our pulsed gate analysis could be extended to this architecture, we focus here on standard linear traps.

\subsection{Coupling neighbouring ions} \label{sec:neighb}

We are interested in the scaling performance of pulsed gate schemes, as well as the laser requirements for particular benchmarks.
We explore the performance of fast gates applied to the first two ions of a chain, as well as the effect of performing the gates on the central ion pair.

\begin{SCfigure}[][t]
	\centering
	\includegraphics[width=0.5\columnwidth]{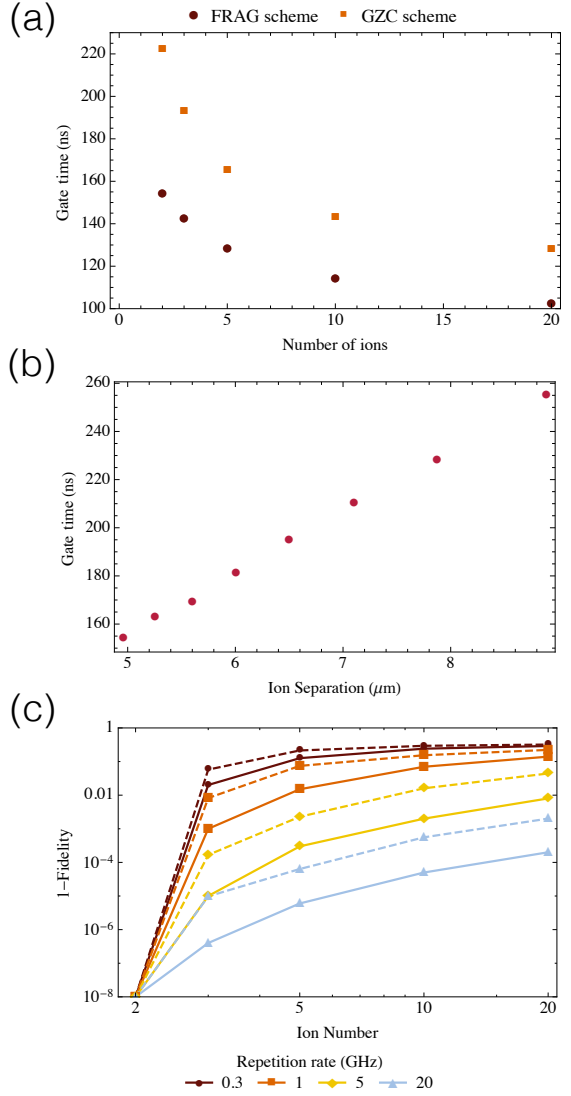} 
	\caption{Inter-ion distance has a strong effect on gate time and fidelity.
	(a) Gate time for fixed trap frequency $\nu = 1.2 \times 2\pi$~MHz as a function of ion number in the trap.
   		The gate is performed between the first two ions in the crystal, with a 5GHz repetition rate.
   		The fidelity is above 0.99 for all save the 10 and 20 ion GZC data points, with fidelity 0.98 and 0.96 respectively.
   	(b) Gate time for FRAG scheme as a function of the distance between two trapped ions.
   		The same number of pulses are used for each gate.
		The gate time depends linearly on ion separation.
	(c) FRAG (solid line) and GZC (dashed line) scheme error as a function of ion number, for ions added to a trap with fixed axial confinement $\nu$.
		The schemes are applied with an optimised number of pulses for each repetition rate and ion number.
	} 
	\label{fig:distfigs}
\end{SCfigure}

When ions are added to a trap with frequency $\nu$, with the frequency fixed and independent of the number of ions, the inter-ion spacing decreases.
Increasing the number of ions in a trap thus increases the coupling strength of the target ions, and the gate is performed faster as shown in Figure~\ref{fig:distfigs}(a).

The infidelity grows with the number of ions in a trap with fixed frequency, as seen in Figure \ref{fig:distfigs}(c).
Although added ions reduce the gate time, the motional requirements on the gate time for high fidelity also become more stringent.
Here the number of pulses in the applied gate is chosen to provide the maximal fidelity, given the repetition rate and number of ions.
Faster repetition rates give faster gates with higher fidelity, using more pulses.

As shown in Figures~7(a) and 7(c), the GZC and FRAG schemes respond similarly as ions are added to the trap.
However, the FRAG scheme is both faster and achieves higher fidelities.
For the following results, we consider just the FRAG scheme for simplicity due to these advantages.

Varying the trap frequency changes the separation of ions in a trap.
In Figure~\ref{fig:distfigs}(b), we show the gate time in terms of the separation distance of two ions.
Here we have a fixed number of pulses $N$, and to reach the necessary relative phase, the phase equation~(\ref{eq:pphasecond}) determines that the inverse cubic dependence of the acquired phase on the ion separation is counteracted by a linear increase in pulse timings.
The gate time thus scales linearly with the inter-ion distance, as shown in the figure.

\begin{figure}[t]
	\centering
	\includegraphics[width=\columnwidth]{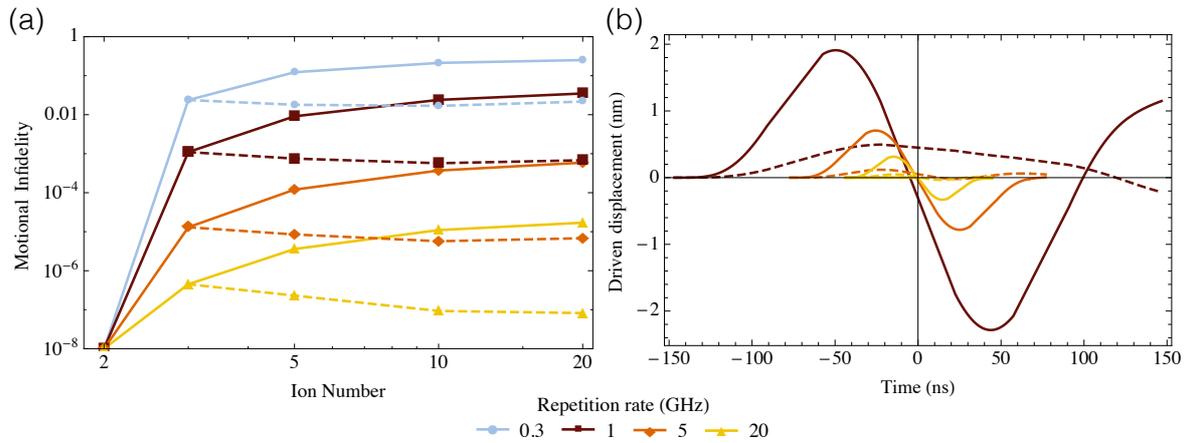} 
	\caption{The distance between addressed ions is fixed, to clarify the ion scaling behaviour.
	(a) Motional infidelity as a function of ion number given a repetition rate. 
		The gate takes 494~ns, 293~ns, 154~ns, and 89~ns for repetition rates of 0.3~GHz, 1~GHz, 5~GHz and 20~GHz respectively.
		Solid lines are for the gate performance on the first two ions in the crystal, dashed lines represent the gate performance on the middle ions.
		(b) Driven displacement from free evolution of ions three (solid) and four (dashed) over the gate operation with different repetition rates, for 5 ions in the crystal.
		The 0.3~GHz repetition rate is neglected for clarity.
		Increasing the repetition rate reduces the displacement of neighbouring ions during and after the operation.
		The local oscillation period of the third ion in the crystal $T_3$, neighbouring the target ions, is around 280ns.
		For repetition rates of 1, 5 and 20~GHz respectively, the gate times are $1.05 T_3$, $0.55 T_3$ and $0.32 T_3$, reflecting the stability gained by operating faster than this timescale.
		Both addressed ions are in the excited state.
	}
	\label{fig:fixeddscal}
\end{figure}

The distance scaling obscures the effect of adding ions to the crystal.
As ions are added, we can relax the trap frequency to maintain a constant distance, and coupling strength, between two ions.
An optimal fast gate applied to these two ions has pulse timings that vary minimally with the number of ions in the crystal; the gate is ion-number independent.
In Figure~\ref{fig:fixeddscal}, we use this method to apply an optimal fast gate to the first two ions in the crystal, with different numbers of total ions.
Fidelity decreases with the crystal length as the motion of each ion is not completely restored.
While the distance between ions one and two is fixed, the distance between ions two and three decreases as more ions are added to the crystal, as can be observed from the ion crystal equilibrium positions in Figure~\ref{fig:eqpos}.
The ions neighbouring the two target ions thus become increasingly coupled to the motion of the target ions as the crystal grows in length.

We similarly fix the distance of the middle two ions in the crystal and apply the fast gate to these two ions, also shown in Figure~\ref{fig:fixeddscal}(a).
Here the fixed distance between the middle, target, ions is smaller than the distance from a middle ion to its outer neighbour.
The fidelity increase in the figure is caused by this decrease in motional coupling to neighbouring ions.
This difference is more pronounced for even numbers of ions, such that there truly are two middle ions in the crystal.
As the number of ions increases, the middle ions and their neighbours in the crystal become approximately equidistant, and the motional fidelity reaches an asymptote.
For gates between middle or end ions in the crystal, the repetition rate dependence is clear; higher fidelities correspond to faster gate times.

The effect of gate speed on motional stability is clearly seen in Figure~\ref{fig:fixeddscal}(b).
The driven displacement of the local ions is restored effectively at the end of the faster gates, as the motional restoration equations reach their stable regime.
This same behaviour is seen when the gate is performed on the middle ions of the crystal rather than the first two ions.
As shown in \cite{Zhu06EL}, for gate times faster than the ion recoil frequency, only local phonon modes are excited and only neighbouring ions affect the gate operation. 
In Figure \ref{fig:fixeddscal}(b), there are five ions in the crystal and the local oscillation timescale of the third ion is 280ns.
The gate times for the system, corresponding to different repetition rates, pass below this timescale and show correspondingly stable results in (a) and (b) of the figure. 
We will consider the displacement amplitude and the harmonic approximation in section~\ref{sec:slims}.

\clearpage

\subsection{Arbitrary couplings}

Now that we have explored the laser regimes for high-fidelity, fast gates between neighbouring ions in a chain, we consider non-neighbouring ions.
It is possible to perform scalable operations using only neighbouring-ion operations and SWAP gates to couple distant ions.
However, it would certainly be simpler if a direct entanglement operation between distant ions were achievable.
We consider the laser requirements of pulsed fast gates between two distant ions.

First, consider the mechanism for coupling neighbouring ions in a large crystal.
High fidelity gates involve fast gate times such that only local modes are involved in the gate operation.
In this regime, there can be no communication between distant ions via shared modes.

The phase equation~(\ref{eq:pphasecond}) quantifies the interaction strength and the gate time required for the requisite relative phase.
For distant ions, the coupling coefficients decay with the cube of the separation.
Either a longer entanglement timescale or larger momentum transfers are required to excite non-local modes.
The motional condition equations~(\ref{eq:pmotcond}) don't change depending on the ions addressed, meaning that sufficiently fast gates will restore every mode, where the required timescale depends on the scheme stability as discussed.

\begin{SCfigure}[][t]
   \centering
   \includegraphics[width=0.5\columnwidth]{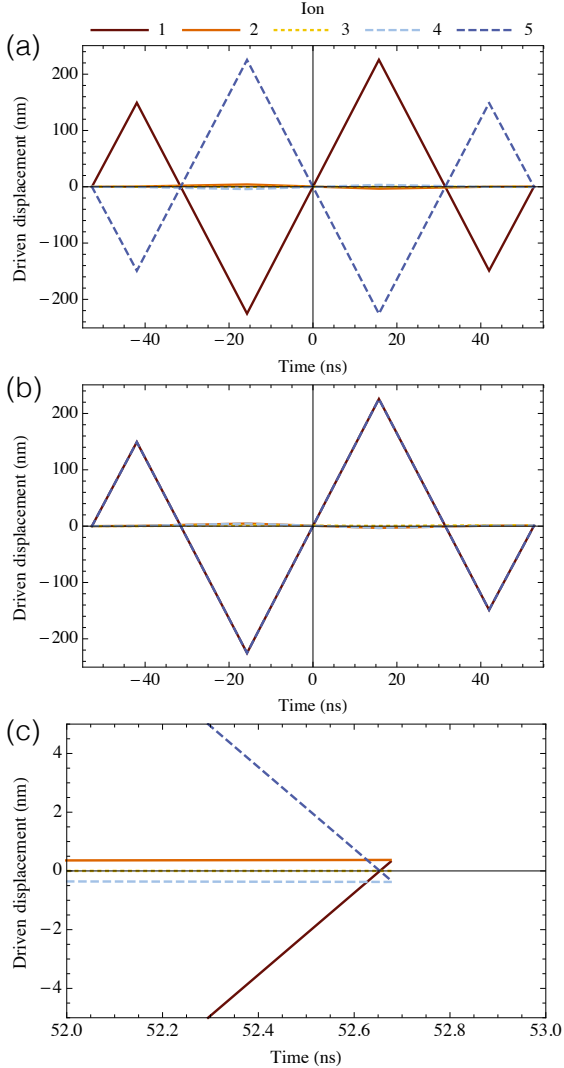}  
   \caption{(a) Driven displacement from free evolution of the five ions in the trap during the FRAG controlled phase gate operation.
   Ions one and five are entangled by the operation, with fidelity of 0.99 using large instantaneous momentum transfers with $n=400$.
   Here ions one and five are in the same internal state (both ground or excited), and we consider equal and opposite momentum kicks for each state for clarity in this displacement figure.
   (b) Ions one and five have different internal states for these dynamics, $|eg\rangle$ or $|ge\rangle$.
   (c) Ions one and five have the same internal states for these dynamics.
   The final displacement of the ions from their initial positions is much less than 1~nm.
    }
   \label{5ions}
\end{SCfigure}

We approach the problem by first finding the optimal solutions to the instantaneous momentum kick case, with an infinite laser repetition rate.
The effect of a non-local gate on the ion crystal can be seen in Figure~\ref{5ions}, where a FRAG gate entangles ions one and five in a five-ion crystal.
Intermediate ions are key to the coupling of the distant ions, and accordingly have state-dependent motion.
The driven motion of ions one and five is much greater than the driven displacement of the intermediary ions.
To couple the ions with fidelity $\geq 0.99$, very large instantaneous momentum kicks are applied; $n=400$, thus the largest kick uses 800 $\pi$-pulse pairs.
Evolution between the momentum kicks appears linear, showing that the free evolution of the ions dominates.
In Figure \ref{5ions}(c), the restoration of the ions' motion is shown to be effective for each ion.

Our analysis of the GZC and FRAG solutions in the two-ion and neighbouring-ion cases, which used finite laser repetition rates, revealed that the fidelity was preserved remarkably well with fast repetition rates, at quantum computing threshold levels.
We now apply finite repetition rates for our distant-ion gates.

\begin{SCfigure}[][t]
   \centering
   \includegraphics[width=0.6\columnwidth]{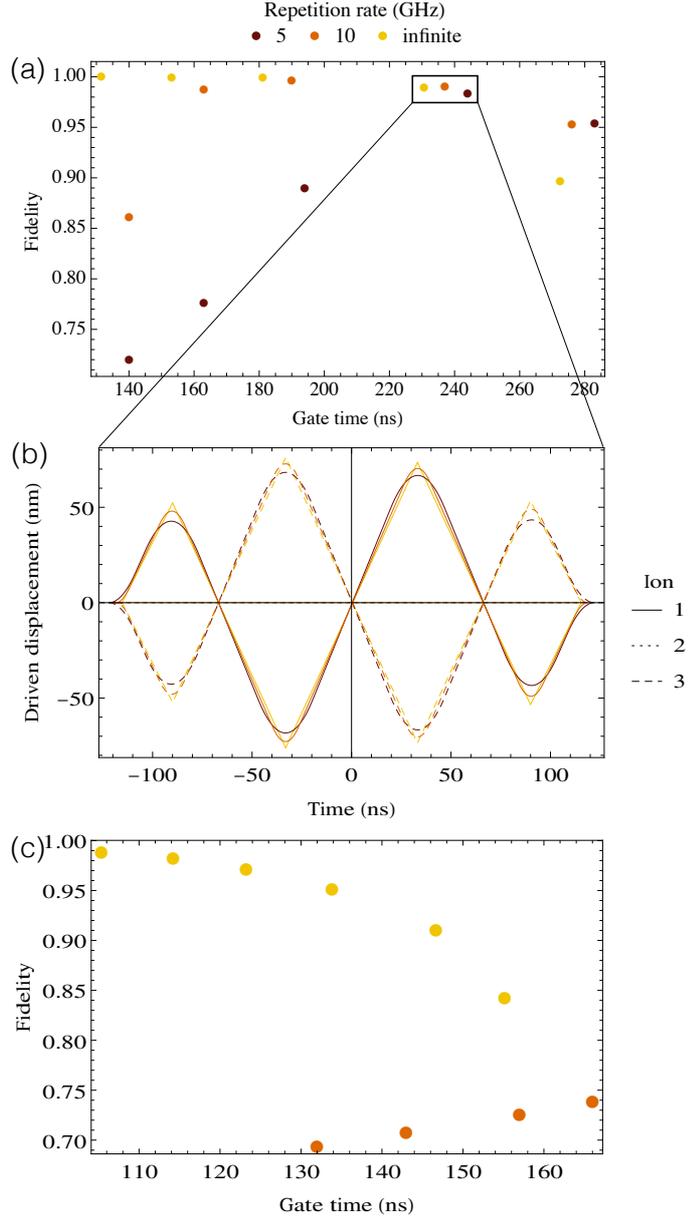}  
   \caption{(a) Optimal fidelity for a given gate time and repetition rate using the FRAG scheme to couple ions one and three in a three ion crystal.
   The three comparable data points marked by a rectangle are explored in (b).
   (b) Driven displacement from free evolution of the three ions in the trap during the controlled phase gate operation.
   The different lines correspond to the different repetition rates of (a), for the marked optimal schemes with gate time close to 240ns, and fidelity close to 0.98.
   Ions one and three are in the excited and ground states respectively, with equal and opposite momentum kicks applied to the states for clarity of displacements.
   (c) Optimal fidelities for entangling ions one and five of a five ion crystal.
    }
   \label{3ions}
\end{SCfigure}

Schemes with finite repetition rates have a maximum momentum transfer in time, unlike the instantaneous pulse scheme of Figure~\ref{5ions}.
This can be seen, in the form of a maximum curvature, for the driven displacement of the ions.
Figure \ref{3ions}(b) shows this effect in a displacement plot of three ions, where ions one and three are coupled, for three different repetition rates.
The schemes are optimal given the repetition rates and gate time $\sim 140$~ns, and have comparable fidelities close to 0.98.
The slowest repetition rate, 5~GHz, requires the largest momentum transfer to reach the necessary phase-space area corresponding to the target relative phase, given its lower curvature limited by pulse separation time.

Figure~\ref{3ions}(a) shows the fast gate fidelity as a function of gate time, coupling ions one and three in a three-ion crystal.
The infinite repetition rate corresponding to instantaneous large momentum kicks gives fidelity arbitrarily close to 1 as the gate time decreases and the motional fidelity of the scheme improves.
As the gate time increases, the motional instability grows and the fidelity drops away.
Finite repetition rates have poor performance when the gate is not long enough to acquire the target relative phase, as the repetition rate limits the total number of pulses in a given gate time.
The optimal gate performance with a finite repetition rate occurs where the gate is long enough to entangle the distant ions with the desired phase, and short enough that the gate is motionally stable.
Higher repetition rates reach a higher optimal fidelity with faster gate times.
Lower repetition rates spread the applied force over longer time.  
This causes the motional stability to improve with increasing gate time, and be lower overall.

Figure~\ref{3ions}(c) shows the gate fidelity for coupling ions one and five in a five-ion crystal.
The effect of repetition rate is stronger for more ions; a repetition rate of 10~GHz is not sufficient for high fidelities. 
Even with an infinite repetition rate, achieving high fidelity requires much lower gate times than for gates coupling ions one and two or one and three.

\section{Limitations} \label{sec:slims}

The fast timescales of the entangling gates we have explored are achieved by careful control of momentum transfers.
Imperfect $\pi$-pulses lead to incomplete population transfer, leading to unwanted state changes, incomplete phase-space trajectories and motional heating.
The timing of the pulses is also important, and timing errors also affect the final motional state and phase acquisition.
General characterisation of these errors is limited, as they depend on the scheme applied and the number of pulses.
As explored in \cite{Bent13NJP}, the pulse timing is remarkably stable up to systematic shifts of beyond 10ps, and this stability is reflected in the robust nature of the GZC and FRAG schemes under non-instantaneous laser pulses.
In contrast, the pulse area can have a dramatic effect; in \cite{Bent13NJP} the authors found that a 1\% systematic pulse energy error leads to a worst case fidelity estimate of almost 1\% for just a four pulse pair scheme.
The worst case fidelity contains combinatorially increasing prefactors of the error in pulse area with the number of pulses.

Initial motional states have no effect on perfect two-qubit gates, however imperfections or more qubits in the crystal introduce initial state dependence.
The fidelity measure, equation~(\ref{eq:fidmeas}), decays exponentially as the product of the mean vibration mode occupation and the final vibrational displacement error.
This effect is quantified in Figure~\ref{Fidforce}, where the impact of mean motional occupation is shown for imperfect two-qubit gates.
Note also that gates performed in sequence will lead to an increasing error as the small heating from the first gate amplifies the error in the second gate, and so on.

The driven motion of the addressed ions increases for faster gates or more distant ions.  
The lengthscale of this motion is still much smaller than the laser wavelength for the data we have explored, meaning that addressibility is no issue.  
However, the harmonic approximation made for the motional modes breaks as the ion displacement becomes significant relative to the ion separation.
A driven displacement of 100~nm is still much smaller than the $\sim 5~\mu$m ion separation, and indeed oscillation amplitudes on this scale only change the harmonic spacing of the relative motion frequency by $0.2\%$ \cite{Wine98JRNIST}.
Entangling distant ions requires increasing maximal displacements as more intermediary ions are added, and the inter-ion separation also decreases.
As the current limits on laser repetition rates are overcome, increasing the possible momentum transfers, anharmonicities will become a limiting factor.
Typical laser powers required for $^{171}$Yb$^+$ can be estimated using the 12~nJ $\pi$-pulse energy presented in \cite{Mizr13APB}: average laser powers of around 12~W and 60~W correspond to repetition rates of 1~GHz and 5~GHz respectively.

\section{Conclusions}

We have considered a general formalism for scalable fast gates using pairs of $\pi$-pulses.
Proposed pulsed gate schemes were analysed for different laser repetition rates, and the FRAG scheme was found to provide optimal gate times and fidelities.
The relationship between repetition rate, gate time and fidelity was extended to neighbouring and non-neighbouring ions in long crystals.
Increasingly large momentum transfers, corresponding to fast repetition rates, are required as ions are further separated in the crystal both to preserve motional robustness and to provide coupling faster than the local oscillation frequencies of the ions.
We present here the repetition rates required to achieve information processing benchmark fidelities and times for two-qubit gates in arbitrary-length ion crystals.

\section{Acknowledgments}
This work was supported by the Australian Research Council Centre of Excellence for Quantum Computation and Communication Technology (Project number CE110001027) (ARRC), Australian Research Council Future Fellowship (FT120100291) (JJH) as well as DP130101613 (JJH, ARRC).
The authors thank D. Kielpinski for helpful discussions.

\clearpage
\appendix

We present three appendices to consider in detail the effect of symmetry on the motional conditions, the details of deriving our fidelity equations, and the adapted equations for treating ions with momentum kicks applied to only one internal state (`asymmetric' kicks).

\section{Symmetry and the motional conditions} \label{app:motsymm}

Equation~(25) in the main text gives the general condition for restoring motional mode $p$: 
\begin{align}
 C_p = -i \sum^{N}_{k=1} c_{pk} e^{i \nu_p t_k} = 0, \label{apeq:pmotcondfull}
\end{align}
where the final displacement of the motional mode is given by
\begin{align}
C_p &\equiv  -2 i \eta_p (b_1^{(p)} \sigma_1^z + b_2^{(p)} \sigma_2^z) \sum^{N}_{k=1} z_k e^{i \nu_p t_k}, \\
        &= i s_p \sum^{N}_{k=1} z_k e^{i \nu_p t_k} \text{, for}\\
s_p &\equiv -2 \eta_p (b_1^{(p)} \sigma_1^z + b_2^{(p)} \sigma_2^z).
\end{align}

Expanding the exponential as a Taylor series in $C_p$,
\begin{align}
C_p &= i s_p \sum^{N}_{k=1} z_k (1 + i \nu_p t_k + \frac{- \nu_p^2 t_k^2}{2} + \frac{-i \nu_p^3 t_k^3}{6} + ...),
\end{align}
and for $C_p = 0$, we require that
\begin{align}
\text{Re}(C_p) &= s_p \sum^{N}_{k=1} z_k (- \nu_p t_k + \frac{ \nu_p^3 t_k^3}{6} +...)  =0, \\
0 &= \sum^{N}_{k=1} z_k (t_k - \frac{ \nu_p^2 t_k^3}{6} -...), \\
\text{Im}(C_p) &= s_p \sum^{N}_{k=1} z_k (1 - \frac{\nu_p^2 t_k^2}{2} +...)  =0, \label{eq:appimcond} \\
0 &= \sum^{N}_{k=1} z_k (1 - \frac{\nu_p^2 t_k^2}{2} +...).
\end{align}

For $1 \gg \nu_p^2 t_k^2$ for each $t_k$, the motional conditions reduce to
\begin{align}
0 &= \sum^N_{k=1} z_k t_k , \label{eq:mcf1}\\
0 &= \sum^N_{k=1} z_k, \label{eq:mcf2}
\end{align}
with no mode dependence.  
This regime and these conditions correspond to free evolution, before the trap evolution has any significant effect.
The required regime may be difficult to reach since the dominant term in our phase condition contains the third order of the pulse times, indicating that phase accrual has strong dependence on the trap evolution time. 
Strong momentum kicks are required to accommodate shorter gate times, as explored in the main text.
Close to this ideal regime, error in the scheme ($C_p \neq 0$) will be mode dependent, and the dominant error term is from the sum in equation~(\ref{eq:appimcond}), proportional to $z_k \nu_p^2 t_k^2$ for each kick $k$.
We can broaden the mode-independent regime by imposing constraints on the scheme pulse directions and timings.  
This provides increasing motional robustness.

A simple example is to impose a reflected antisymmetry, such that the momentum kicks $\underbar{z}$ and their timings $\underbar{t}$ obey

\begin{center}
\begin{tabular}{cccccccccc}
$\underbar{z}$ &= ($a_1$, &$a_2$, &..., &$a_m$, &$|$ &$-a_m$, &$-a_{m-1}$, &..., &$-a_1$) \\
$\underbar{t}$ &= ($-\tau_1$, &$-\tau_2$, &..., &$-\tau_m$ &$|$ &$\tau_m$, &$\tau_{m-1}$, &..., &$\tau_1$),
\end{tabular}
\end{center}

This sets $\text{Im}(C_p)$ to zero by virtue of its even powers of $t_k$, while we still require
\begin{align}
\text{Re}(C_p) &= 2 s_p \sum^{m}_{k=1} a_k \sin (\nu_p \tau_k) = 0, \\
0 &= \sum^{m}_{k=1} a_k \tau_k - a_k \frac{\nu_p^2 \tau_k^3}{6} + ....
\end{align}

The highest order term in our new motional condition equation corresponds to equation (\ref{eq:mcf1}), while we have subsumed equation (\ref{eq:mcf2}) into the symmetry condition.
Assuming mode independence means that the dominant error term is proportional to $a_k \nu_p^3 \tau_k^3$, which is on the order of the cube of the pulse times.
The asymmetry decreases the error size from a quadratic power of pulse times above, providing motional stability.

We can introduce a more complex symmetry to further expand our region of mode-independent total gate times:

\begin{center} 
\begin{tabular}{llllllll} \label{app:symm2}
$\underbar{z}$ = ($a_1$, &..., $a_l$, &$|$ $-a_l$, &..., $-a_1$, &$|$ $-a_1$, &..., $-a_l$, &$|$ $a_l$, &..., $a_1$) \\
$\underbar{t}$ = ($-f-\tau_1$, &..., $-f-\tau_l$, &$|$ $-f+\tau_l$, &..., $-f+\tau_1$, &$|$ $f-\tau_1$, &..., $f-\tau_l$, &$|$ $f+ \tau_l$, &..., $f+\tau_1$),
\end{tabular}
\end{center}
for $f>\tau_1>\tau_2>...>\tau_l$.
Here $\text{Re}(C_p)$ is zero, and we require
\begin{align}
\text{Im}(C_p) &= 2 s_p \sum^{l}_{k=1} a_k \cos (\nu_p (f+\tau_k)) - a_k \cos (\nu_p (f-\tau_k)) = 0, \\
0 &= \sum^{l}_{k=1} a_k \left[ \nu_p^2 2 f \tau_k - \frac{\nu_p^4}{24} ((f+\tau_k)^4 - (f-\tau_k)^4) + ...   \right] ,
\end{align}
where if the first term dominates the sum, we have a mode-independent condition equivalent to equation (\ref{eq:mcf1}).
The largest error term in the sum is on the order of $a_k \nu_p^4 f^3 \tau_k$, on the fourth order of the pulse times for $f \sim \tau_1$.
Again, the added symmetry conditions increase the motional stability region for the scheme.

Adding ions reduces the mode-independent regime, since the mode frequency plays a part in the terms we require to be negligible, and these frequencies increase with ions added.
Extra degrees of freedom found in the GZC and FRAG schemes relative to the Duan single degree of freedom also provide extra motional stability.
Note also that while motional conditions can be satisfied independent of ion numbers for sufficiently fast gates, the phase equation still has mode dependence.
However, only local phonon modes can be excited as discussed in the main text.

\section{Fidelity measure derivation} \label{supp:fid}

The computational fidelity before state averaging is given by
\begin{align}
F_1 = \text{Tr}_m[\langle \psi_0| U_{\text{id}}^\dagger U_{\text{re}}| \psi_0 \rangle \langle \psi_0 | \otimes \rho_m U_{\text{re}}^\dagger U_{\text{id}} |\psi_0\rangle],
\end{align}
where $U_{\text{re}}$ and $U_{\text{id}}$ denote the real and ideal gate operations, including motional displacement for the real gate as performed by our gate schemes to follow.  The motional trace $\text{Tr}_m$ is taken along with the inner product of the ideal and real operations with respect to the internal ion states.  The initial internal state of the ions is $|\psi_0\rangle$, and the initial motional state is given by the density operator $\rho_m$.  In the computational basis $\{ |gg\rangle, |ge\rangle, |eg\rangle, |ee\rangle  \}$, the ideal unitary is given by
\begin{align}
U_I = \left( \begin{array}{cccc}
e^{i \pi/4} & 0 & 0 & 0 \\
0 & e^{-i \pi/4} & 0 & 0 \\
0 & 0 & e^{-i \pi/4} & 0 \\
0 & 0 & 0 & e^{i \pi/4} \end{array} \right),
\end{align}
up to global phase. 

The real operator, representing our approximation to the ideal unitary with some error, is given by
\begin{align}
U_{\text{re}} = \left( \begin{array}{cccc}
e^{i \phi_{gg}} \hat{D}_{gg} & 0 & 0 & 0 \\
0 & e^{i \phi_{ge}} \hat{D}_{ge} & 0 & 0 \\
0 & 0 & e^{i \phi_{eg}} \hat{D}_{eg} & 0 \\
0 & 0 & 0 & e^{i \phi_{ee}} \hat{D}_{ee} \end{array} \right),
\end{align}
representing phase changes $\phi_{ij}$ as well as motional displacements $\hat{D}_{ij}$ specific to internal states, $i,j \in \{g,e \}$.  
Note that there are no off-diagonal terms as the initial internal states are preserved through the real operation due to the assumed perfect $\pi$-pulse pairs providing momentum kicks while perfectly restoring state population.

Recall that the final displacement is given in equation~(\ref{apeq:pmotcondfull}) for each mode $p$:  
\begin{align}
C_p \equiv -i \sum^N_{k=1} c_{pk} e^{i \nu_p t_k}.
\end{align}
The total displacement operator is the tensor product of each mode displacement.
The internal state dependence comes from $c_{pk}$:  
\begin{align}
c_{pk} 	&= 2 z_k (b_1^{(p)} \sigma_1^z + b_2^{(p)} \sigma_2^z) \eta_p, \\
C_p 		&= -2 i (b_1^{(p)} \sigma_1^z + b_2^{(p)} \sigma_2^z) \eta_p \sum^N_{k=1} z_k e^{i \nu_p t_k}. \label{eq:cpfid}
\end{align}
We have defined the motional operators
\begin{align}
\hat{D}^{(p)}_{ij} 	&= \langle i_1 j_2 | \hat{D}_p (C_p) |i_1 j_2\rangle, \label{eq:Dijdef}
\end{align}
for $i$ and $j$ internal states of ions one and two.
Each mode displacement is related; for both ions in the same or opposite internal states these relations are simple:
\begin{align}
\hat{D}^{(p)}_{gg} &= (\hat{D}^{(p)}_{ee})^\dagger \\
\hat{D}^{(p)}_{ge} &= (\hat{D}^{(p)}_{eg})^\dagger
\end{align}
due to the opposite direction of each displacement kick.
The other relations are by a scalar displacement amount, such that the motional displacement $C_p$ for internal states $|ee \rangle$ and $| eg \rangle$ are related by
\begin{align}
C_p (ee) = \frac{b_1^{(p)}  + b_2^{(p)} }{b_1^{(p)} - b_2^{(p)} } C_p (eg).
\end{align}
The tensor product composite, mode independent, displacement operators for each state are related in the same way:
\begin{align}
\hat{D}_{gg} &= \Pi_{p=1}^L \hat{D}^{(p)}_{gg} = \hat{D}_{ee}^\dagger \\
\hat{D}_{ge} &= \Pi_{p=1}^L \hat{D}^{(p)}_{ge} = \hat{D}_{eg}^\dagger.
\end{align}

We derive the phase acquired for each state from equation~(24) in the main text, the mode-dependent phase term: 
\begin{align}
\xi_p = \sum^N_{m=1} \sum^{m-1}_{k=1} c_{pm} c_{pk} \sin (\nu_p (t_m - t_k) ) - \Re [ \alpha \sum^N_{k=1} c_{pk} e^{-i \nu_p t_k} ],
\end{align}
for an initial motional coherent state $\alpha$.
This second component of the phase is zero when the motion is restored at the end of the gate.
As discussed in the main text, when the displacement is nonzero and this term is significant, we can correct for it using single-ion rotations as it is not entangling.
The total phase term, from equation~(30) in the main text neglecting global phase, reduces to: 
\begin{align}
\sum_{p=1}^L \xi_p = 8 \sum_{p=1}^L \eta_p^2 \sigma_1^z \sigma_2^z b_1^{(p)} b_2^{(p)} \sum^N_{m=1} \sum^{m-1}_{k=1} z_{m} z_{k} \sin (\nu_p (t_m - t_k) ).
\end{align}
The presence of both $U_{\text{re}}$ and its Hermitian conjugate in the fidelity equation ensure that global phase does cancel.
Internal states determine the value of the sum, and we have defined 
\begin{align}
\phi_{ij} = \langle i_1 j_2 | \sum_{p=1}^L \xi_p |i_1 j_2 \rangle,
\end{align}
for $i,j \in \{g,e\}$.

Total phase acquired by particular internal states is related according to
\begin{align}
\phi_{gg} = -\phi_{ge} = -\phi_{eg} = \phi_{ee},
\end{align}
as each mode-dependent term in the sum changes sign for different internal states of ions one and two.

For a general initial internal state
\begin{align}
|\psi_0 \rangle &= a_{00}|gg\rangle + a_{01} |ge\rangle + a_{10} |eg\rangle + a_{11} |ee\rangle, \\
A 	&= \langle \psi_0| U_{\text{id}}^\dagger U_{\text{re}} |\psi_0 \rangle \\
	&= |a_{00}|^2 e^{i(\phi_{gg}-\pi/4)} \hat{D}_{gg} + |a_{01}|^2 e^{i(\phi_{ge}+\pi/4)} \hat{D}_{ge} + |a_{10}|^2 e^{i(\phi_{eg}+\pi/4)} \hat{D}_{eg} \nonumber \\
	& \quad + |a_{11}|^2 e^{i(\phi_{ee}-\pi/4)} \hat{D}_{ee}.
\end{align}
Using the cyclic nature of the trace, we have
\begin{align}
F_1 &= \text{Tr}_m [A^\dagger A \rho_m].
\end{align}

The trace becomes an expectation of pairs of motional displacement operators with respect to $\rho_m$.  
Since operators on different motional modes commute, we can group the mode-dependent components, for example
\begin{align}
\hat{D}_{gg} \hat{D}_{ge} &= \Pi_{p=1}^L \hat{D}^{(p)}_{gg} \hat{D}^{(p)}_{ge}.
\end{align}
For a single mode, products of displacement operators are determined by
\begin{align}
\hat{D}(a)\hat{D}(b) &= e^{(ab^* - a^*b)/2} \hat{D}(a+b) \\
&= \hat{D}(a+b) \text{  if } b=ca \text{ for a scalar c.}
\end{align}
We have seen that the arguments of our displacement operators, $C_p$ from equation~(\ref{eq:Dijdef}), are indeed related by scalars for the same mode and different internal states.

If the initial motional state $\rho_m$ is separable with respect to the motional modes, then the expectation values can be taken individually for separate modes.  
We assume an initial thermal product state as a typical motional distribution, 
\begin{align}
\rho_m^{(p)} &= (1 - e^{-\hbar \nu_p /k T}) \sum_{n=0}^\infty  |n \rangle \langle n| e^{-n \hbar \nu_p /k T}  \\ 
\rho_m &= \rho_m^{(1)} \otimes \rho_m^{(2)} \otimes ... \otimes \rho_m^{(L)},
\end{align}
represented in the number basis, with temperature $T$ and $k$ the Boltzmann constant. 

The expectation value for a single mode with a displacement $z$ follows:  
\begin{align}
\langle \hat{D} (z) \rangle_{\rho_m} 
	&= (1 - e^{-\hbar \nu_p /k T}) \sum_{n=0}^\infty \langle n | \hat{D}_p (z) |n \rangle e^{-n \hbar \nu_p /k T} \\
	&= e^{-|z|^2 (1/2 + \bar{n}_p)},
\end{align}
where $\bar{n}_p$ is the average phonon population of the $p$th motional mode.

Choosing a particular internal state has limited use as a figure of merit for a quantum gate, which takes inputs of different initial forms in practice, the fidelity of each being important.  
We consider the average fidelity, taking an even weighting over all initial states.
This corresponds to integrating over $F_1$ for the possible values of the coefficients; an integral over the three-dimensional surface (given four basis states) of a hypersphere of radius one. 
We now have the formalism in place to calculate the average fidelity, which depends on the number of ions as this determines the phase and displacement equations.  
For two ions,
\begin{align}
F_{2\text{ave}} = \frac{1}{12} \left( 6 + e^{-4 m_1 |C_1|^2} + e^{-4 m_2 |C_2|^2} + 4 e^{- (m_1 |C_1|^2 + m_2 |C_2|^2)} \sin (2 \phi_{gg})  \right) ,
\end{align}
where $m_i \equiv (1/2 + \bar{n}_i)$ is a function of the motional mode mean occupation level, and $C_p$ is the final displacement for mode $p$ defined in equation~(\ref{eq:cpfid}) using equal and opposite internal states for nonzero $C_1$ and $C_2$ respectively.

For more ions, there are more complicated internal state dependencies in the $C_p$ final displacements, so we define
\begin{align}
c_p = 2 \eta_p \sum^N_{k=1} z_k e^{i \nu_p t_k},
\end{align}
which is independent of the internal state for a simpler fidelity definition. 
For three ions, the fidelity becomes
\begin{align}
F_{3\text{ave}} &= \frac{1}{12} (6 + e^{-2 m_2 |c_2|^2 -6 m_3 |c_3|^2} + e^{-5.\bar{33} m_1 |c_1|^2 - 2 m_2 |c_2|^2 - 0.\bar{66} m_3 |c_3|^2} \nonumber \\
& \quad + 2(e^{-1.\bar{33} m_1 |c_1|^2 - 2.\bar{66} m_3 |c_3|^2 } + e^{-1.\bar{33} m_1 |c_1|^2 - 2 m_2 |c_2|^2 - 0.\bar{66} m_3 |c_3|^2 }  ) \sin{2 \phi_{gg}}  ).
\end{align}
The average fidelity for no relative phase and very large motional displacement is 0.5, a point to keep in mind as we consider the fidelity values for the schemes in the main text.

\section{Asymmetric momentum kick equations} \label{app:asymm}

If the laser providing the momentum kicks addresses a transition from one computational state to an auxiliary state, then the other computational state is untouched by the `kick'.  Without loss of generality, we choose the ground state to undergo the momentum kick, while the excited state is left invariant.  The unitary kick operator, equation~(12) in the main text, becomes
\begin{align}
U_{kick} = e^{-2izk(x_1 \alpha_1^z + x_2 \alpha_2^z)} \label{eq:ukickas},
\end{align}
where $\alpha_i^z |0\rangle_i = 0$ and $\alpha_i^z |1\rangle_i = |1\rangle_i$.
Thus $(\alpha_i^z)^2 = \alpha_i^z$ and $\alpha_i^z = (I_i + \sigma_i^z)/2$ for identity operator $I$.

The mode basis transformation and the displacement and rotation operator expansion take place as for symmetric kicks in section~4.1 of the main article, with the adapted motional displacements
\begin{align}
C_p &= -2 i \eta_p (b_1^{(p)} \alpha_1^z + b_2^{(p)} \alpha_2^z) \sum^{N}_{k=1} z_k   e^{i \nu_p t_k} \\ 
\end{align}
for each mode.
The phase acquired for a given scheme becomes
\begin{align}
\sum^L_{p=1} \xi_p &= 4 \sum^L_{p=1} \eta_p^2 (b_1^{(p)} \alpha_1^z + b_2^{(p)} \alpha_2^z)^2  \sum^N_{m=2} \sum^{m-1}_{k=1} z_m z_k \sin (\nu_p (t_m - t_k)) \\
	&= 4 \sum^L_{p=1} \eta_p^2 ((b_1^{(p)})^2 \alpha_1^z + (b_2^{(p)})^2 \alpha_2^z + 2 b_1^{(p)} b_2^{(p)} \alpha_1^z \alpha_2^z)  \sum^N_{m=2} \sum^{m-1}_{k=1} z_m z_k \sin (\nu_p (t_m - t_k)), 
\end{align}
which includes non-entangling state-dependent phase.
This causes deviation from our target gate, which can be corrected by single-qubit gates. 

We thus have the asymmetric phase condition equation
\begin{align}
\frac{\pi}{4} = 2 \sum^L_{p=1} \eta_p^2 b_1^{(p)} b_2^{(p)} \sum^N_{m=2} \sum^{m-1}_{k=1} z_m z_k \sin(\nu_p(t_m-t_k)), \label{eq:asphase}
\end{align}
where the right hand side is one quarter the size of the symmetric kick phase equation~(32).
A symmetric scheme that solves the condition equations will thus provide only $\pi/16$ phase in equation~(\ref{eq:asphase}), or $\pi/8$ relative phase between symmetric ($|00\rangle$, $|11\rangle$) and asymmetric ($|10\rangle$, $|01\rangle$) states. 
The adapted displacements and phase can be directly substituted into the fidelity derivation.

\clearpage

\bibliographystyle{ieeetr} 
\bibliography{libraryallr}

\end{document}